\documentclass[twoside]{dis04}

\usepackage{picins}
\def\runauthor{Stefan Gieseke}
\def\shorttitle{Herwig++}
\def\ee{e^+e^-}
\def\qq{q\bar q}
\def\qqg{q\bar qg}
\def\HW{\textsc{\small HERWIG}}
\def\pythia{\textsc{\small PYTHIA}}
\def\hpp{\mbox{\textsf{Herwig++}}}

\def\Pythia7{\textsf{Pythia7}}
\def\SHERPA{\textsf{SHERPA}}

\def\ycut{y_{\rm cut}}

\begin{document}

\title{The new Monte Carlo Event Generator \hpp{}}


\author{Stefan Gieseke}
\address{University of Cambridge, Cavendish Laboratory,\\
  Madingley Road, Cambridge CB3\,0HE, United Kingdom}

\maketitle

\abstracts{We present results obtained with the new Monte Carlo event
  generator \hpp{}.  In its first version (1.0), \hpp{} is capable of
  simulating $\ee$ Annihilation events.  We compare results for
  different distributions with a vast amount of available data from
  $\ee$ annihilation.}

\section{Introduction}
The advent of the future Large Hadron Collider (LHC) requires new
tools for the Monte Carlo (MC) simulation of final states.  As the
well--established tools, e.g.\ \pythia{} \cite{Pythia} and \HW{}
\cite{Herwig64} have grown tremendously during the LEP era and are
considered difficult to maintain and even more difficult to extend, it
was decided to rewrite them in an object oriented language, in
particular C++.  Therefore, the old codes are presently being
rewritten under the names \Pythia7{} \cite{Pythia7} and \hpp{}
\cite{Herwig++}.  Furthermore, a new project with the same goals,
\SHERPA{} \cite{SHERPA}, has been established.  In this talk we
consider our own project \hpp{}.  This program is available in version
1.0 and capable of simulating $\ee$ annihilation events, particularly
at LEP energies.  We compare our results to event shape data and
exclusive particle spectra mainly taken at the LEP experiments but
also by SLD and JADE.

\section{Simulation}

\begin{figure}[!p]
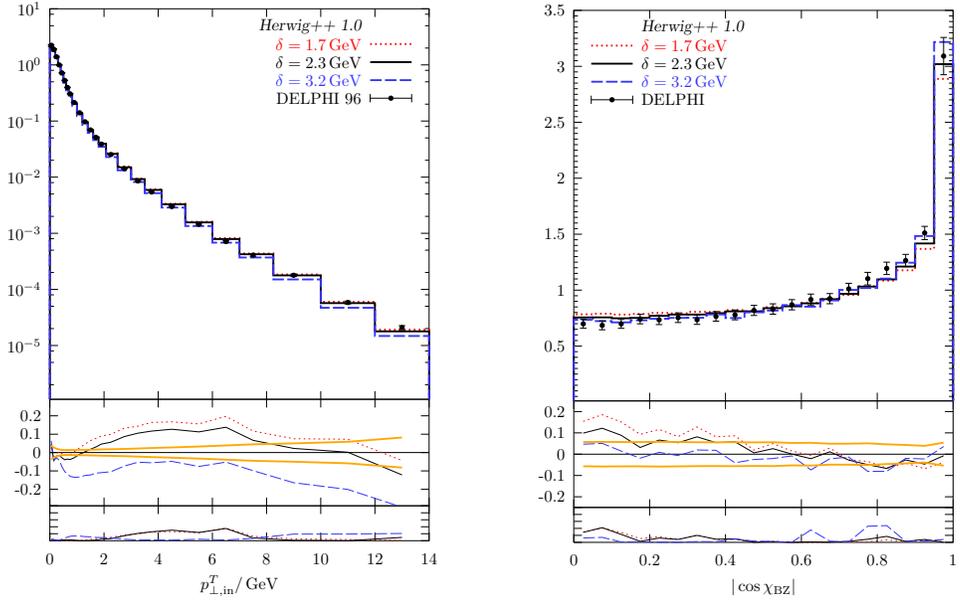

\begin{center}
\epsfig{file=es.33, scale=0.62}
\hfill
\epsfig{file=es.71, scale=0.62}
\end{center}
\caption{The distribution of transverse momentum in the event plane
  w.r.t.\ the thrust axis (left) and of the Bengtsson--Zerwas angle
  (right). \label{fig:ptintbz}}
\end{figure}

The main steps of the physical simulation of an $\ee$ annihilation
consist of a hard subprocess, a parton shower evolution, cluster
hadronization and hadronic decays.  In \hpp{}~1.0 we have a basic
$\ee\to(\gamma, Z)\to\qq$ tree level matrix element (ME).  In
addition, we apply ME corrections of two different natures in order to
be able to match up the phase space space accessible by the parton
shower with that of the full matrix element at the lowest order for
$\qq g$ production.  For the hard ME correction \cite{MikeMEC}, one
additional gluon is emitted according to the full ME whenever a trial
emission would hit the so-called dead region in the $\qq g$ phase
space (cf.\ Fig.~\ref{fig:dead}).  For the soft ME correction we
consider the parton shower emission from, say, the quark (cf.\ region
$Q$ in Fig.~\ref{fig:dead}).  Comparing the expression for one parton
shower emission with that for the full $\qq g$ matrix element we can
compute an additional weight or apply an extra veto in order to
distribute parton shower emissions closer to the actual matrix element
in this phase space.  Applying the soft correction to the 'hardest
emission so far' we continue the parton shower evolution until a lower
cutoff scale is reached ($\sim \delta$ in
Figs.~\ref{fig:njetsthrust},~\ref{fig:ptintbz}).  We carry out the
parton shower evolution in terms of a set of new evolution variables
as described in \cite{NewVariables}.

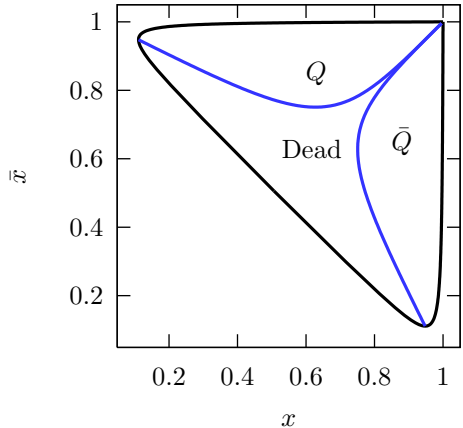
\piccaption[]{The phase space for $\qqg$ production in $\ee$
  annihilation for heavy quarks. Using the new parton shower variables
  the regions $Q$ and $\bar Q$ are populated by the first parton
  shower emission whereas the 'Dead' region is populated with the help
  of a hard matrix element correction.\label{fig:dead}}
\parpic(6cm,6cm)[r]{
  \epsfig{file=3jps.2} 
}
\piccaption[]{}
These allow a smooth coverage of soft gluon phase space (cf.\ 
Fig.~\ref{fig:dead}, $x, \bar x\to 1$) and a correct treatment of
heavy quarks in the quasi collinear limit. Next, due to
preconfinement, the low scale partons are paired up into colourless
clusters which still carry the flavour information of their
constituting quarks.  In this way we will find a cluster mass spectrum
which is invariant against variations of the initial cm energy of the
$\ee$ annihilation.  Clusters which still have a large invariant mass
are fissioned and finally decayed into hadrons according to some
phenomenological weights.  Here, we have introduced a different
interpretation of the diquark weight in order to have a more
consistent baryon production mechanism \cite{Herwig++}, cf.\ also
\cite{kupco}.  Finally, all unstable hadrons decay into stable ones,
predominantly pions and hard photons.

\section{Results for \boldmath $\ee$ Annihilations}

\begin{figure}[!p]
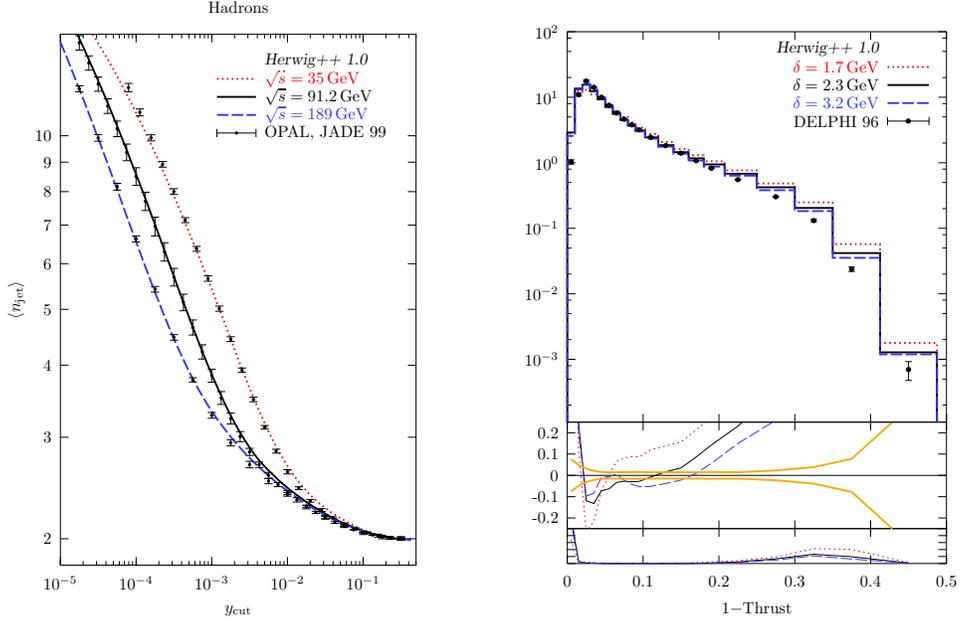

\begin{center}
\epsfig{file=njets2.7, scale=0.62}
\hfill
\epsfig{file=es.1, scale=0.62}
\end{center}
\caption{The multiplicity of minijets ($k_\perp$ algorithm) at PETRA,
  LEP and LEPII energies from \hpp{} in comparison with data (left).
  The thrust distribution compared to DELPHI data
  (right).\label{fig:njetsthrust}}
\end{figure}

We used the program \hpp{} to simulate a large number of observables
in $\ee$ collisions and compared the results to LEP \cite{LEPdata} and
SLD data as well as older data from PETRA \cite{Petradata}.  We
covered the available spectrum of data ranging from exclusive hadron
multiplicities and momentum distributions over all kinds of event
shape distribution to four jet angles.  We wanted to give a reasonable
overall description of the available data without emphasising any
particular observable too much and came up with a 'recommended'
parameter set for \hpp{} \cite{Herwig++}.  Here, we want to focus on a
few examples.  First, we adjusted the hadronization parameters to
reasonable values in considering multiplicity data.  In
Fig.~\ref{fig:njetsthrust} (left) we consider the number of minijets
in the $k_\perp$--algorithm at different collider energies and find a
very good description of the data.  Despite the fact that we apply
matrix element corrections, the thrust distribution tends to overshoot
for too spherical and too point--like events.  Quite in contrast, we
find that the distribution of transverse momentum in the thrust event
plane is described very well.  As we haven't attempted a detailed
tuning of our generator, we assume that these minor contradictions
will be properly adjusted in the future.  Considering four jet events
at $\ycut = 0.08$ we find that our description of e.g.\ the
Bengtsson--Zerwas angle distribution is very good.  This is by no
means guaranteed as we have not used any four--jet matrix element in
our code.  In addition, we have been able to give a very good
description of the $B$--fragmentation function (for a figure, see
\cite{Mrinaltalk}).  This exemplifies the quality of our new parton
shower for heavy quark physics as we can simply tune the parton shower
accordingly, attributing heavy quark partonic effects to the
perturbative domain as one might expect.

We conclude that we have succeeded in delivering the new Monte Carlo
event generator \hpp{} which is perfectly capable of describing LEP
physics in a similar or better quality as it was known from \HW{}.
This is considered a first step towards a new generator that will be
suitable for the description of hadronic events at HERA, the Tevatron
and the LHC.


\section*{Acknowledgements} I would like to thank the other \hpp{} 
authors for a very pleasant and fruitful collaboration.  I wish to
thank the organisers for a very pleasant working atmosphere and
particularly F.\ Krauss for some interesting discussions.

\end{document}